\documentstyle[12pt,epsfig]{article}
\begin{document}

\begin{titlepage}
\rightline{September 2009}
\vskip 2cm
\centerline{\Large \bf  
Evidence for mirror dark matter from}
\vskip 0.4cm
\centerline{\Large \bf  
the CDMS low energy electron recoil spectrum}

\vskip 2.2cm
\centerline{R. Foot\footnote{
E-mail address: rfoot@unimelb.edu.au}}

\vskip 0.7cm
\centerline{\it School of Physics,}
\centerline{\it University of Melbourne,}
\centerline{\it Victoria 3010 Australia}
\vskip 2cm
\noindent
We point out that mirror dark matter predicts low 
energy ($E_R \stackrel{<}{\sim} 2$ keV) electron
recoils from mirror electron scattering
as well as nuclear recoils from mirror ion scattering.
The former effect is examined and applied to the recently released low energy
electron recoil data from the CDMS collaboration. 
We speculate that the sharp
rise in electron recoils seen in CDMS below 2 keV might be due to mirror
electron scattering and show that the parameters suggested by the data
are roughly consistent with the mirror dark matter explanation of the
annual modulation signal observed in the DAMA/Libra and DAMA/NaI experiments.  
Thus, the CDMS data offer tentative evidence supporting the mirror
dark matter explanation of the DAMA experiments, which can be
more rigorously checked by future low energy electron recoil
measurements.

\end{titlepage}

A successful dark matter theory emerges if one 
postulates that the fundamental laws governing the interactions of the dark matter
sector are identical to the ordinary matter sector.
That is, we assume the existence of a sector which is an exact duplicate of the 
ordinary matter sector, so that
the fundamental Lagrangian is:
\begin{eqnarray}
{\cal L} = {\cal L}_{SM} (e, u, d, \gamma,...) + {\cal L}_{SM} (e', u', d', \gamma', ...)
\end{eqnarray}
Such a dark matter theory can be motivated from simplicity and
minimality.
It can also be motivated from a symmetry reason if left and right handed chiral
fields are interchanged in the extra sector. This is because the theory then
exhibits the space-time $Z_2$ parity symmetry, $x \to -x$, where chiral left (right) handed ordinary
fermions transform into chiral right (left) handed mirror fermions; and ordinary bosonic fields
transform into mirror bosonic fields (see ref.\cite{flv} for the precise transformation).
Since this theory enlarges the space-time symmetry to include parity (as well as the other
improper Lorentz transformations), we refer to the particles 
in the extra sector as mirror particles. The standard model extended with
a mirror sector was first studied in ref.\cite{flv} and shown to be a phenomenologically consistent 
renormalizable theory
(for a review and more complete list of references see ref.\cite{review})
\footnote{
Note that successful big bang nucleosynthesis (BBN) and 
large scale structure (LSS) requires effectively asymmetric initial
conditions in the early Universe, $T' \ll T$ and $n_{b'}/n_b \approx 5$. 
See ref.\cite{some} for further discussions.}.

If we include all interaction terms consistent with renormalizability and the symmetries of
the theory then we must add to the Lagrangian a $U(1)$ kinetic mixing
interaction\cite{fh} and Higgs - mirror Higgs quartic coupling\cite{flv}:
\begin{eqnarray}
{\cal L}_{mix} = \frac{\epsilon}{2} F^{\mu \nu} F'_{\mu \nu} + 
\lambda \phi^{\dagger}\phi \phi'^{\dagger}\phi' \ ,
\end{eqnarray}
where $F_{\mu \nu}$ ($F'_{\mu \nu}$) is the ordinary (mirror) $U(1)$ gauge boson 
field strength tensor and
$\phi$ ($\phi'$) is the electroweak Higgs (mirror Higgs) field.
The most general Higgs potential, including the quartic Higgs mixing term (above)
was studied in ref.\cite{flv} and shown to have 
the vacuum $\langle \phi \rangle = \langle \phi' \rangle$ for
a large range of parameters. With this vacuum, the masses
of the mirror particles are all identical to their ordinary
matter counterparts. Note that interactions between ordinary and mirror particles
at low energies are expected to be dominated by the $U(1)$ kinetic
mixing interaction. This interaction leads to photon - mirror photon kinetic
mixing, which provides the key to testing this theory (for a review,
see e.g. ref.\cite{review}).

In this framework, dark matter is comprised of stable massive mirror particles: 
$e', H', He', O'...$ etc, with known masses. Dark matter
in the galactic halo is then a spherically distributed self interacting mirror 
particle plasma which can be stable from collapse
provided that a heat source(s) exists. In fact, ordinary supernova
can plausibly supply the required heating if photon-mirror photon
are kinetically mixed with $\epsilon \sim 10^{-9}$\cite{sph}
\footnote{ 
A mirror sector with such kinetic mixing is consistent with
all known laboratory, astrophysical and 
cosmological constraints\cite{sasha, sil, paolo, lab, rafelt}
.}.
For kinetic mixing of this magnitude 
about half of the total energy emitted in ordinary Type II Supernova
explosions ($\sim 3\times 10^{53}$ erg) will be in the form of light mirror 
particles ($\nu'_{e,\mu,\tau}$, $e'^{\pm}, \gamma'$)
originating from kinetic mixing induced plasmon decay into $e'^+ e'^-$ in the 
supernova core\cite{rafelt}.
This implies a heating of the halo (principally due to the $e'^{\pm}$ component), of around:
\begin{eqnarray}
L^{SN}_{heat-in} \sim \frac{1}{2} \times 3\times 10^{53}\ erg 
{1 \over 100\ years} \sim 10^{44}\ {\rm erg/s \ for \ Milky\ Way
}
\end{eqnarray}
It turns out that this matches (to within uncertainties) the energy lost from the halo due to
radiative cooling\cite{sph}:
\begin{eqnarray}
L^{halo}_{energy-out} = \Lambda \int_{R_1} n^2_{e'} 4\pi r^2 dr \sim 10^{44} \ {\rm erg/s \ for
\ Milky \ Way}.
\end{eqnarray}
In other words, a gaseous mirror particle halo can potentially survive without collapsing because
the energy lost due to dissipative interactions can be replaced by the energy from ordinary supernova
explosions. Presumably there is some detailed dynamical reasons maintaining this balance,
which of course, may be difficult to elucidate due to the complexity of that particular problem.

Kinetic mixing of around $\epsilon \sim 10^{-9}$ is also implicated\cite{fdama} by the annual modulation
signal found in the DAMA/NaI and DAMA/Libra experiments\cite{dama}.
In ref.\cite{fdama} (updating and improving earlier studies\cite{fdama2}), it was shown
that elastic scattering of the heavy mirror oxygen ($O'$) component
off the target nuclei
can nicely explain the DAMA signal if $\epsilon \sim 10^{-9}$ including the 
observed recoil energy dependence of the annual modulation amplitude. Importantly,
this explanation is also consistent
with the null results of the other direct detection experiments.

In view of the success of the mirror dark matter theory in explaining
the existing direct detection experiments,
it is important to look for other ways to further test this theory. 
In this note we would like to point out another signature of mirror dark matter
relevant for direct dark matter detection
experiments, which has not been previously discussed.
In addition to target nuclear recoils, low energy electron recoils are also predicted
arising from mirror electron ordinary electron scattering. 
In order
to compute the mirror electron interaction rate,
we need to know the number density of mirror electron's at the Earth's location, their
velocity distribution, and interaction cross section.

Assuming that the dark matter density is\footnote{We use natural
units $\hbar = c = 1$ unless otherwise stated.}  $\omega = 0.3\ GeV/cm^3$ at the earth's location,
and that the halo mass is dominated by the $H', He'$ component [with mass fraction 
$Y_{He'} \equiv n_{He'} m_{He}/(n_{H'}m_H + n_{He'}m_{He})$],
we expect a mirror electron number density, $n_{e'}$, of:
\begin{eqnarray}
n_{e'} = {\omega
\over m_p} \left( 1 - \frac{Y_{He'}}{2} \right)
\end{eqnarray}
where $m_p$ is the proton mass.

The halo mirror particles are presumed to form a self interacting 
spherically distributed plasma at temperature $T$.
[This is necessary to explain the flat rotation curves in spiral galaxies].
The dynamics of the mirror particle plasma has been investigated
previously\cite{sph,fdama2},
where it was found that the condition of hydrostatic equilibrium implied that the
temperature of the plasma satisfied:
\begin{eqnarray}
T = {1 \over 2} \bar m v_{rot}^2 \ ,
\label{4}
\end{eqnarray}
where $\bar m = \sum n_i m_i/\sum n_i$ [$i=e', H', He', O'....$] is the mean mass of 
the particles in the plasma, and $v_{rot} \approx 254$ km/s is the rotational velocity of the sun
around the center of the galaxy\cite{rot}.
Assuming the plasma is completely ionized, a reasonable approximation since in turns out that
the temperature of the plasma is $\approx \frac{1}{2}$ keV (see below) we find:
\begin{eqnarray}
{\bar m \over m_p} = {1 \over 2 - \frac{5}{4} Y_{He'}} 
\end{eqnarray}
Hence from Eq.(\ref{4}) we find the temperature to be:
\begin{eqnarray}
T \approx 0.48 \ keV \ {\rm for \ Y_{He'} = 1} \nonumber \\
T \approx 0.18 \ keV \ {\rm for \ Y_{He'} = 0}
\end{eqnarray}
Thus the temperature is quite sensitive to $Y_{He'}$, but mirror BBN calculations
suggest $Y_{He'} \approx 0.9$ for $\epsilon \sim 10^{-9}\cite{bbn}$.
Anyway, the gas of mirror particles will have a Maxwellian velocity distribution,
\begin{eqnarray}
f_i (v) &=& e^{-\frac{1}{2} m_i v^2/T}  \nonumber \\  
 &=& e^{-v^2/v_0^2[i]} 
\end{eqnarray}
where the index $i$  labels the particle type [$i=e', H', He', O', ...$].

Clearly, the velocity dispersion of the particles in the mirror matter halo depends
on the particular particle species and from Eq.(\ref{4}) satisfies:
\begin{eqnarray}
v_0^2 [i] = v_{rot}^2 \frac{\overline{m}}{m_i}
\end{eqnarray}
Note that if $m_i \gg \overline{m}$, then $v_0^2[i] \ll v_{rot}^2$. Consequently heavier 
mirror particles have their velocities (and hence energies) relative to the earth
boosted by the Earth's (mean) rotational velocity
around the galactic center, $v_{rot}$. 
This allows a heavy mirror particle in the mass range $\sim m_0 \sim 15$ GeV 
to provide a significant annual modulation signal in the
energy region probed by DAMA ($2 < E_R^m/keV < 6$), which it turns out
has the right properties to fully account\cite{fdama} for the data presented
by the DAMA collaboration\cite{dama}.
On the other hand the mirror electron, being much lighter than the mean mass, $\bar m$,
will have a very large velocity dispersion:
\begin{eqnarray}
v_0^2 (e') = v_{rot}^2 {4m_p \over 3m_e} \ \ {\rm for \ He' \ dominated \ halo, \ i.e. \ Y_{He'} = 1}
\end{eqnarray}
That is, we expect $v_0 (e') \approx 12,000 $ km/s for $Y_{He'} \approx
1$. Note that at these velocities, we
can, to a good approximation, neglect the motion of the Earth through the halo as far
as the $e'$ component is concerned.
Thus, since $T \stackrel{<}{\sim} \frac{1}{2} keV$, we expect an electron scattering 
signal to be confined to low
energy electron recoils $\stackrel{<}{\sim}$ 2 keV.

We now discuss the interaction cross section between ordinary and mirror
electrons.
The effect of the photon-mirror photon kinetic mixing is to induce a small
coupling of ordinary photons to mirror electrons, of magnitude $\epsilon e$.
This enables mirror electrons to Rutherford scatter off free ordinary electrons with
cross section in the non-relativistic limit:
\begin{eqnarray}
{d\sigma \over dE_R} = {\lambda \over E_R^2 v^2}
\label{cs}
\end{eqnarray}
where 
\begin{eqnarray}
\lambda \equiv {2\pi \epsilon^2 \alpha^2 \over m_e} \
\end{eqnarray}
Here $E_R$ is the recoil energy of the target electron, initially presumed at
rest relative to the incoming mirror electron  of velocity $v$.
This cross section should also approximate the inelastic scattering of halo mirror electrons off
bound atomic electrons provided that
the recoil energies are much larger than the binding energy.

Recently, the CDMS collaboration has released data on low energy electron scattering
down to a recoil energy of $1$ keV on a Germanium target\cite{cdms1,cdms2}, which
we reproduce in figure 1. Such an 
experiment is potentially sensitive to
mirror electron-ordinary electron scattering, where we expect a sharp rise in interaction
rate below $2$ keV. Interestingly, such a rise in interaction rate was seen
in that experiment, although it should also be noted that this
rise is near the CDMS energy threshold.
Certainly, any conclusions
drawn from an analysis of the $E_R < 2$ keV data are tentative and will need to be confirmed
by future low threshold electron recoil measurements.

\vskip 1cm

\centerline{\epsfig{file=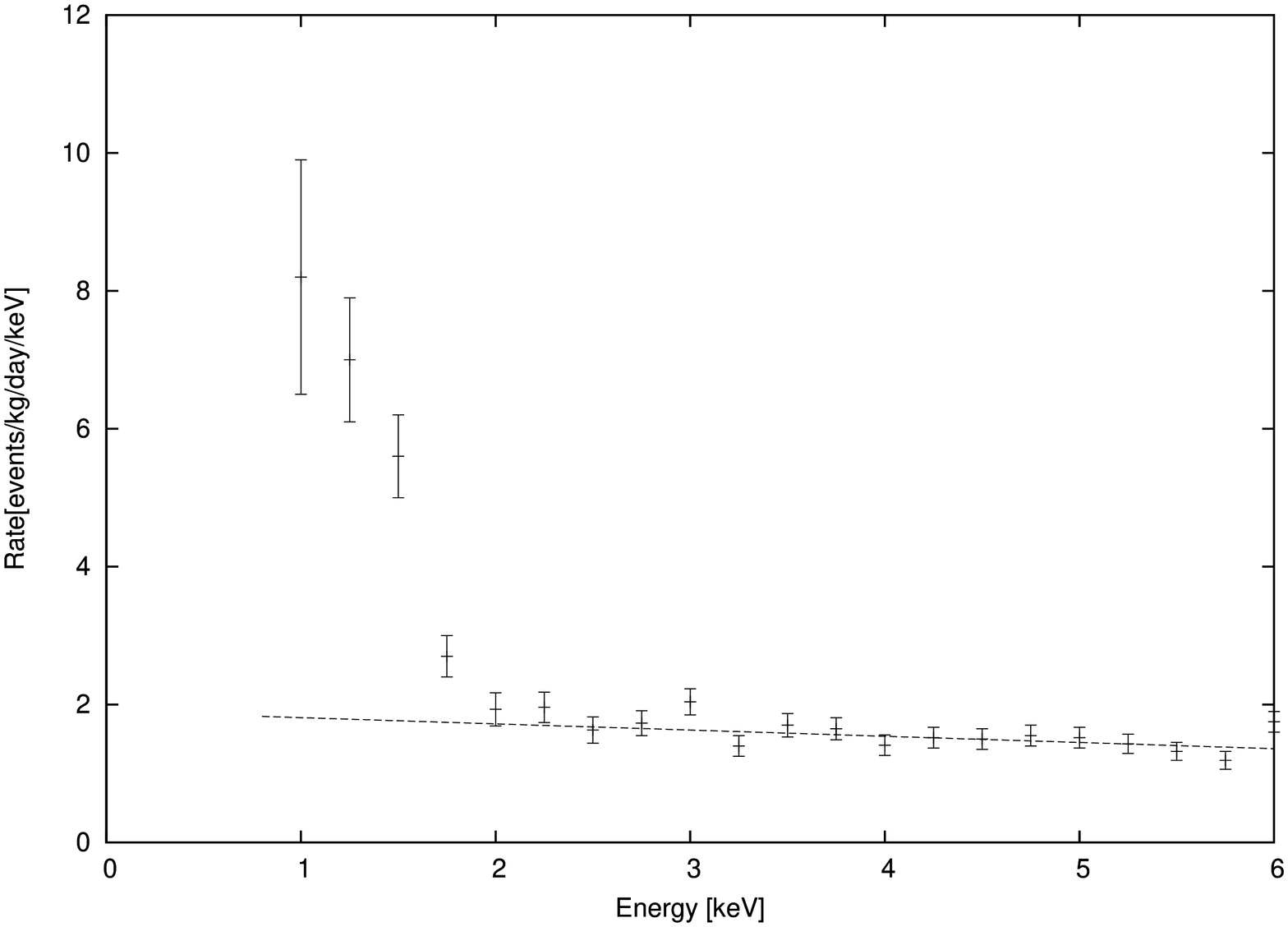,angle=270,width=13cm}}
\vskip 0.5cm

\noindent
Figure 1: Efficiency corrected low-energy electron-scattering spectrum
as measured by CDMS/Ge\cite{cdms1}.
The dashed line is a simple background model.

\vskip 1cm

In principle, it would be desirable to properly account for
the cross section of mirror electrons on bound atomic electrons which
is a non-trivial quantum mechanical problem and is beyond
the scope of this initial study.
In this paper, we will perform a crude analysis using the following 
simple approximation.
We consider the scattering only on the loosely bound Ge electrons 
in the 3d, 4s and 4p outer shells, which have atomic 
binding energy $\stackrel{<}{\sim}$ 30 eV\cite{webelem}. There are 14 such electrons per Ge atom.
The remaining 18 electrons 
occupy the  more tightly bound inner shells all of which have binding energy $\stackrel{>}{\sim} 120$ eV
(and the 10 inner most electrons have binding energy $\stackrel{>}{\sim} 1.2$ keV)
and have typically much faster velocities, comparable to $v_0 (e')$. We expect the interactions 
on these inner shell electrons to be suppressed relative to the more weakly bound outer shell electrons.
Furthermore, we approximate the 14 weakly bound outer shell electrons of Ge, as free and
at rest. We expect the resulting cross section to be valid to within around $20-30\% $ for the
recoil energy range $1 < E_R/keV < 2$.
With this approximation, the predicted differential interaction rate is:
\begin{eqnarray}
{dR \over dE_R} &=& 
gN_T  n_{e'} \int {d\sigma \over dE_R}
{f_{e'}(v) \over k} |v|
d^3v \nonumber \\
&=& g N_T  n_{e'}
{\lambda \over E_R^2 } \int^{\infty}_{|v| > v_{min}
(E_R)} {f_{e'}(v) \over k|v|} d^3 v 
\label{55}
\end{eqnarray}
where $N_T$ is the number of target atoms per kg of detector and 
$k = [\pi v_0^2 (e')]^{3/2}$ is the Maxwellian distribution normalization factor.
The quantity $g = 14$, is the number of loosely bound atomic electrons in Ge 
as we discussed above.
Note that the lower velocity limit,
$v_{min} (E_R)$, 
is given by the kinematic relation:
\begin{eqnarray}
v_{min} &=& \sqrt{ {2E_R\over m_e } } .
\label{v}
\end{eqnarray}
The velocity integral in Eq.(\ref{55}) can be analytically solved leading to:
\begin{eqnarray}
{dR \over dE_R} &=& 
g N_T n_{e'} {\lambda \over E_R^2}
\left( {2e^{-x^2}
\over \sqrt{\pi} v_0 (e') }\right) 
\end{eqnarray}
where $x = v_{min}/v_0$.

In order to compare with the experimentally measured rate, we must
convolve this rate, with a Gaussian to take into account the finite detector
resolution:
\begin{eqnarray}
{dR \over dE_R^m} &=& 
{1 \over \sigma \sqrt{2\pi}} 
\int {dR \over dE_R} e^{-(E_R - E_R^m)^2/2\sigma^2}  \ dE_R
\label{bla}
\end{eqnarray}
Here $E_R^m$ is the `measured recoil energy' while actual recoil energy we 
denote as $E_R$. The
detector averaged resolution, $\sigma$, is measured to 
be\cite{cdms2}: 
\begin{eqnarray}
\sigma = \sqrt{(0.293)^2 + (0.056)^2 E_R}\ keV
\ .
\end{eqnarray}
We also incorporate a detection efficiency effect as follows.
From figure 2 of ref.\cite{cdms2} it is evident that the detection efficiency
goes to zero at $E_R \approx 0.8$ keV, and this suggests a lower recoil energy
limit of around $0.8$ keV. That is, we integrate Eq.(\ref{bla}) 
down to actual recoils of $0.8$ keV. Aside from this, it is unnecessary to include
the detection efficiency explicitly in Eq.(\ref{bla}) since the data presented is efficiency 
corrected.

Numerically, we can now compare the predicted rate described above
with the low energy 
recoil electron scattering data from the CDMS/Ge experiment. The rate depends on the parameters
$\epsilon, Y_{He'}$. We fix $Y_{He'} \approx 0.9$ which is the value predicted from mirror
BBN analysis\cite{bbn}. In figure 2 we give our results for $\epsilon = 6\times 10^{-10}$
(dashed line),
$\epsilon = 7\times 10^{-10}$ (solid line) and $\epsilon = 8\times 10^{-10}$ (dotted line).
We have assumed a nearly flat background rate of R(background) $= 1.9 - 0.09 \times E_R^m$ (shown
as the dashed line in figure 1), which is 
similar to the approach of ref.\cite{cdms1}.

\vskip 1cm
\centerline{\epsfig{file=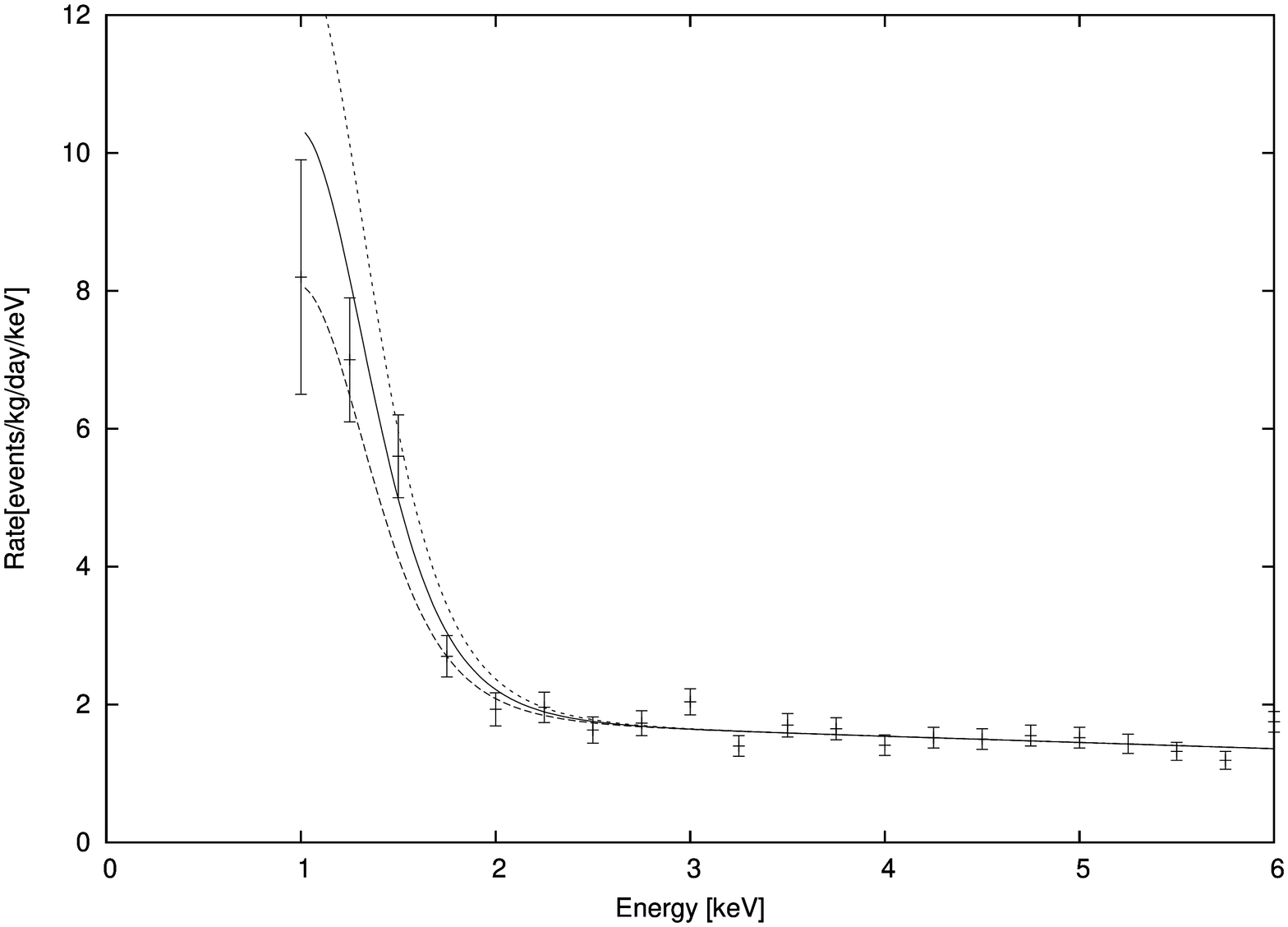,angle=270,width=13cm}}
\vskip 0.5cm
\noindent
Figure 2: Mirror dark matter induced electron recoils + background model, for
the parameters $Y_{He'} = 0.9$ and $\epsilon = 6\times 10^{-10}$ (dashed line),
$\epsilon = 7\times 10^{-10}$ (solid line) and $\epsilon = 8\times 10^{-10}$ 
(dotted line), compared with the CDMS/Ge data.

\vskip 1cm

Anyway,
despite the various systematic uncertainties and other deficiencies in our analysis,
it is nevertheless quite interesting that
the value of $\epsilon$ suggested by the data 
is around $\epsilon \approx 7 \times 10^{-10}$.
This value is consistent with the mirror dark matter explanation of the DAMA annual modulation
signal which
requires:
\begin{eqnarray}
\epsilon \sqrt{{\xi_{O'} \over 0.1}}
\approx 10^{-9}
\end{eqnarray}
where $\xi_{O'} \equiv n_{O'} m_{O}/\omega$ is the proportion of $O'$ by mass 
in the galactic halo. If taken seriously the CDMS data suggest a relatively
large mirror metal ($O'$) proportion in the halo. Such a scenario is possible
given the asymmetric evolution of ordinary and mirror matter implied by the
asymmetric initial condition $T' \ll T$ inferred to exist in the early Universe (see
e.g. ref.\cite{some} for discussions).  
For example
$T' \ll T$ in the early Universe implies a
higher $Y_{He'}$ fraction (c.f $Y_{He}$) 
which leads to more rapid mirror star evolution and early structure formation
in the mirror sector\cite{mstar}. The more rapid stellar evolution in the mirror sector
might be expected to lead to a higher metal proportion c.f. the ordinary
matter sector.

Finally in figure 3 we examine the effect of changing the $Y_{He'}$ parameter for fixed $\epsilon$.
Fixing $\epsilon = 7\times 10^{-10}$, we show results for $Y_{He'} = 0.3, 0.6, 0.9$.
As the figure demonstrates the results dependent somewhat sensitively on the value of $Y_{He'}$
which is due to the Temperature dependence (or equivalently, $v_0 (e')$ dependence) on $Y_{He'}$
as discussed earlier.
\vskip 1cm
\centerline{\epsfig{file=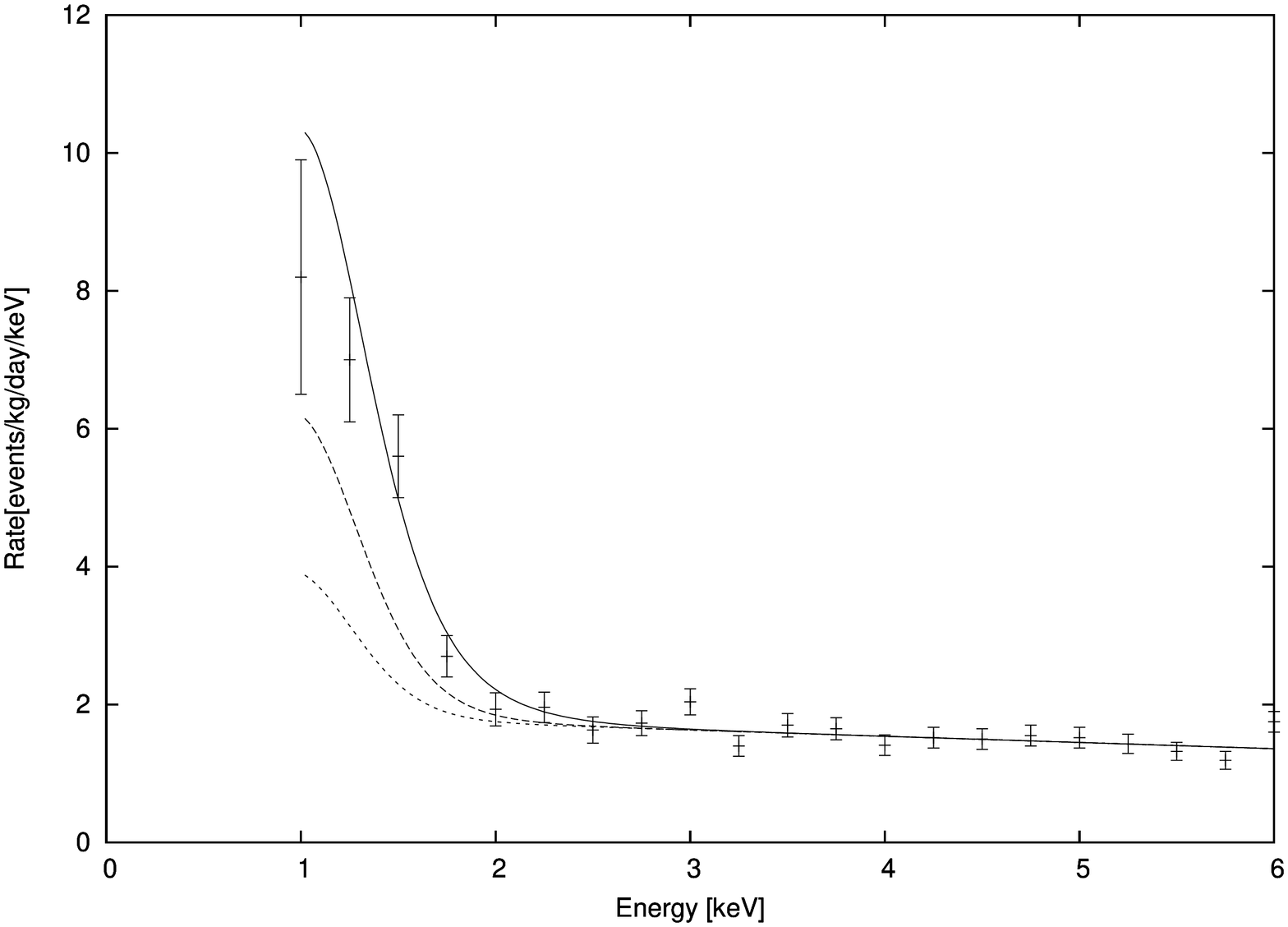,angle=270,width=13cm}}
\vskip 0.5cm
\noindent
Figure 3: Mirror dark matter induced electron recoils + background model, for
fixed $\epsilon = 7\times 10^{-10}$ and $Y_{He'} = 0.3$ (dotted line),
$Y_{He'} = 0.6$ (dashed line) and $Y_{He'} = 0.9$ (solid line).
\vskip 1cm
Recall that the mirror dark matter interpretation of the DAMA annual modulation signal depends
less sensitively on $Y_{He'}$\cite{fdama}. Thus, ultimately low recoil energy electron scattering 
measurements might provide a useful way to measure $Y_{He'}$, which can
then be checked against the theoretical expectation from mirror BBN analysis, which
suggest $Y_{He'} \approx 0.9$\cite{bbn}.

In conclusion, we have 
pointed out that mirror dark matter predicts low energy electron
recoils ($E_R \stackrel{<}{\sim} 2$ keV) from mirror electron scattering
as well as nuclear recoils from mirror ion scattering.
The former effect is examined and applied to the recently released low energy
electron recoil data from the CDMS collaboration. 
We speculate that the sharp
rise in electron recoils seen in CDMS below 2 keV might be due to mirror
electron scattering and show that the parameters suggested by the data
are roughly consistent with the mirror dark matter explanation of the
annual modulation signal observed in the DAMA/Libra and DAMA/NaI experiments.  
This interpretation of the CDMS data can be more rigorously checked by
future low energy electron recoil measurements, which we await with interest.

\vskip 1cm
\noindent 
{\large Acknowledgments}

\vskip 0.2cm
\noindent
This work was supported by the Australian Research Council.

\end{document}